\documentclass[%
 reprint, showpacs,
 amsmath,amssymb,
 aps,
 prl,
]{revtex4-1}

\usepackage{graphicx}
\usepackage{dcolumn}
\usepackage{bm}
\usepackage{subfigure}
\usepackage{color}

\def\be#1\ee{\begin{equation}#1\end{equation}}
\def\bea#1\eea{\begin{align}#1\end{align}}
\def\bse#1\ese{\begin{subequations}#1\end{subequations}}

\def\1#1{{\hat{{\boldsymbol{#1}}}}}                                 		
\def\2#1{\hat{#1}}                                              		   		
\def\3#1{{\mathbf{#1}}}                                             	   		
\def\4#1{{\boldsymbol{#1}}}                                            		
\def\5#1{{\mathcal#1}}                                                            		
\def\6#1{\bar{#1}}                                                           		
\def\7#1{{{#1}}}                                    		
\def\8#1{\widetilde{#1}}
\def\9#1{\check{#1}}
\def\+#1{{\overset{{\scriptscriptstyle +}}{#1}{}}}                  		
\def\b+#1{{\overset{{\scriptscriptstyle +}}{\mathbf{#1}}{}}}        	
\def\g+#1{{\overset{{\scriptscriptstyle +}}{\boldsymbol{#1}}{}}}    

\definecolor{dark-green}{rgb}{0.278,0.7,0.4}                    

\definecolor{my-brown}{rgb}{0.69,0.247,0.13}                    

\definecolor{my-purple}{rgb}{0.47,0.12,0.46}                    

\definecolor{my-greenblue}{rgb}{0.129,0.313,0.419}              

\definecolor{my-orange}{rgb}{1,0.5,0.25}                        

\definecolor{my-red}{rgb}{0.745,0,0.2117}                       

\definecolor{my-gray}{rgb}{0.5,0.5,0.5}                         

\definecolor{my-dark-blue}{rgb}{0.1,0.1,0.7}                    

\definecolor{my-indigo}{rgb}{0.29,0.0,0.51}                    

\begin{document}

\preprint{APS/123-QED}



\title{Beyond the Bode-Fano Bound: Wideband Impedance Matching for Short-Pulses using Temporal Switching of Transmission-Line Parameters}

\author{Amir Shlivinski}%
\affiliation{Department of Electrical and Computer Engineering,
Ben-Gurion University of the Negev, Beer Sheva, Israel, 84105}%
\author{Yakir Hadad}
\email{hadady@eng.tau.ac.il}
\affiliation{School of Electrical Engineering, Tel-Aviv University, Ramat-Aviv, Tel-Aviv, Israel, 69978}
\date{\today}

\begin{abstract}
Impedance matching is one of the most important practice in wave engineering as it enables to maximize the power transfer from the signal source to the load in the wave system. Unfortunately, it is bounded by the Bode-Fano criterion that states, for any passive, linear and time-invariant  matching network,  a stringent tradeoff between the matching-bandwidth and efficiency; implying severe constraints on various electromagnetic and acoustic wave systems.
%
%
Here, we propose a matching paradigm that overcome this issue by  using a temporal switching of the parameters of a metamaterial-based transmission-line, thus
revoking the time-invariance assumption underlying the Bode-Fano criterion.
Using this scheme we show theoretically that an efficient wideband matching, beyond Bode-Fano bound, can be achieved for short-time pulses in challenging cases of very high contrast between the load and the generator impedances, and with significant load dispersion; situations common in e.g., small antennas matching, cloaking, with  applications for ultra-wideband communication, high resolution imaging, and more.

%
%
%
%

\end{abstract}

\pacs{Valid PACS appear here}
\maketitle



\begin{figure}[!h]
\centering
\includegraphics[width=82mm]{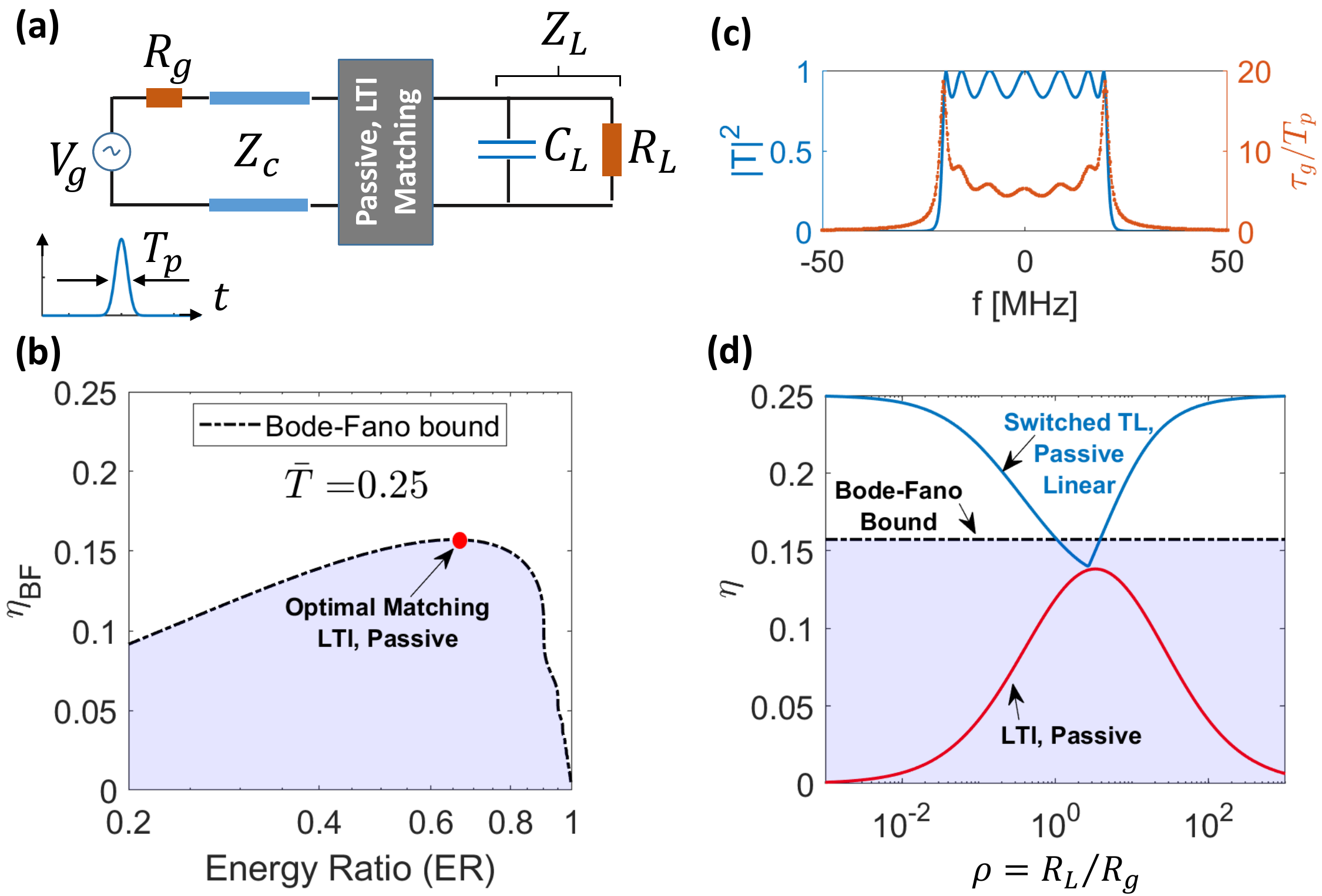}
\caption[]{(a) A generator excites a short-time baseband signal propagating toward a dispersive load with $R_L=5\Omega$, and $C_L=3.14nF$ ($\tau_L=R_LC_L=15.7ns$). The signal bandwidth is BW$=100$MHz$ \rightarrow T_p=62.8ns$. Thus, $\bar T=\tau_L/T_p=0.25<1$, indicating on highly dispersive load \cite[Sec.~5]{SM}. (b) Assuming a passive, LTI matching network, the Bode-Fano bound on the source-load  transmission efficiency $\eta_{\mbox{\scriptsize BF}}$ for the model discussed in (a) is shown as a function of the Energy Ratio (ER) between the excited pulse energy within the matching network passband, and the total excited energy (See \cite[Sec.~5]{SM}). Optimal matching in this case yields $\eta_{\mbox{\scriptsize BF}}\approx0.15$ at $\mbox{ER}\approx0.65$. (c) The transmission and group delay of a nearly optimal, passive, LTI matching network based on a 7'th order Chebyshev design. Its bandwidth, $\sim40$MHz yields the optimal value $\mbox{ER}=0.65$. In its passband, the network is highly dispersive, implying, besides spectrum chopping also an  unavoidable additional distortion of the transmitted signal. (d) A distortion-less, though sub-optimal, matching can be achieved by discarding the matching network in (a) and instead optimizing on $Z_c$ and $R_g$ to maximize energy delivery. The efficiency in this case, as a function of the contrast $\rho=R_L/R_g$ is shown by the red line, which is always below the Bode-Fano limit. However, using the switched TL scheme  discussed in this paper the matching efficiency can significantly exceed the bound as indicated by the blue line and with minimal signal distortion, as discussed below. }
\label{Fig-1}
\end{figure}

\emph{Introduction}.\textemdash
Impedance matching, a fundamental practice in wave engineering \cite{Collin, Kinsler}, is used to maximize the transmission efficiency between a signal generator and a load. However, it is limited by the Bode-Fano criterion \cite{Fano, Bode, Pozar, Monticone, Acher2009} which implies a stringent tradeoff between the matching efficiency and the matching bandwidth in a passive and linear time-invariant (LTI) systems. This has severe implications on system bandwidth, and consequences on signal distortion, particularly in cases of highly dispersive loads, and when the load-generator impedance contrast is high, $Z_g/Z_L\ll1$ or $\gg1$. This is often the case, e.g., in the contexts of matching small antennas for communication and imaging \cite{Balanis}, for cloaking \cite{PaiYen}, radars \cite{DELANEY}, at the acoustic interface between liquid and
gas \cite{Bok2018}, and for ultrasonic absorbers \cite{Acher2009}.
The issue is demonstrated in Fig.~\ref{Fig-1}, where Fig.~\ref{Fig-1}(a) illustrates a typical transmission system, and, Fig.~\ref{Fig-1}(b) shows the Bode-Fano bound calculated for a dispersive load $Z_L$ with parallel $R_L=5\Omega$ and $C_L=3.14nF$, and when the signal source excites a baseband, nearly rectangular pulse, 100MHz in bandwidth. In this case, the maximal transmission efficiency is $\sim15\%$, obtained when the matching frequency passband is set such that it captures $\sim65\%$ of the excited signal power \cite[Sec.~5]{SM}.
%
%
%
%
%
%
%
%
%
%
Moreover, as a consequence of the bound, typically, the matching efficiency, and particularly the transmission group delay, are highly dispersive (see Fig.~\ref{Fig-1}(c)). Implying that a short-time pulse, that propagates in such a system will be severely distorted. Nonetheless, in a passive LTI network, dispersion-less albeit sub-optimal matching, can be obtained for a given load, $Z_L$, by discarding the `matching network' in Fig.~\ref{Fig-1}(a) and instead optimizing the transmission efficiency, $\eta$, over the  transmission-line (TL) impedance, $Z_c$, for a given generator impedance, $R_g$.  The resulting efficiency, $\eta$, is shown by the red line in Fig.~\ref{Fig-1}(d) as function of the contrast $\rho=R_L/R_g$ \cite[Sec.~5]{SM}, with optimum $\sim13\%$ when $R_L/R_g\approx3$ \cite{Comm2}.

How can we bypass the Bode-Fano bound? the idea is to revoke its underlying assumptions \cite{Fano, Bode, Acher2009, Pozar}, namely, \emph{passivity, linearity and time-invariance}.
%
%
Indeed, in the past years active matching networks that are implemented by non-Foster components have been successfully explored as a means to overcome the  matching bound. These have been suggested to increase also the bandwidth of small antennas \cite{Sussman-Fort} and cloaking devices \cite{Pai-Yen, Soric}. However, this approach comes with the unavoidable price of increase  in the internal noise of the network \cite{Pozar},  as well as with inherent potential for instability \cite{Stearns2011,Stearns2012}. Moreover, while non-Foster matching can  compensate reactive loads, it cannot be applied to match resistive loads.

In this letter we propose a different venue to mitigate the Bode-Fano bound by violating the time-invariance assumption through temporal switching of the TL parameters. By this scheme, it is possible to achieve \emph{beyond} Bode-Fano transmission efficiency as shown by the blue line in Fig.~\ref{Fig-1}(d), which is a key result of our work.

\begin{figure}[!h]
\centering
\includegraphics[width=82mm]{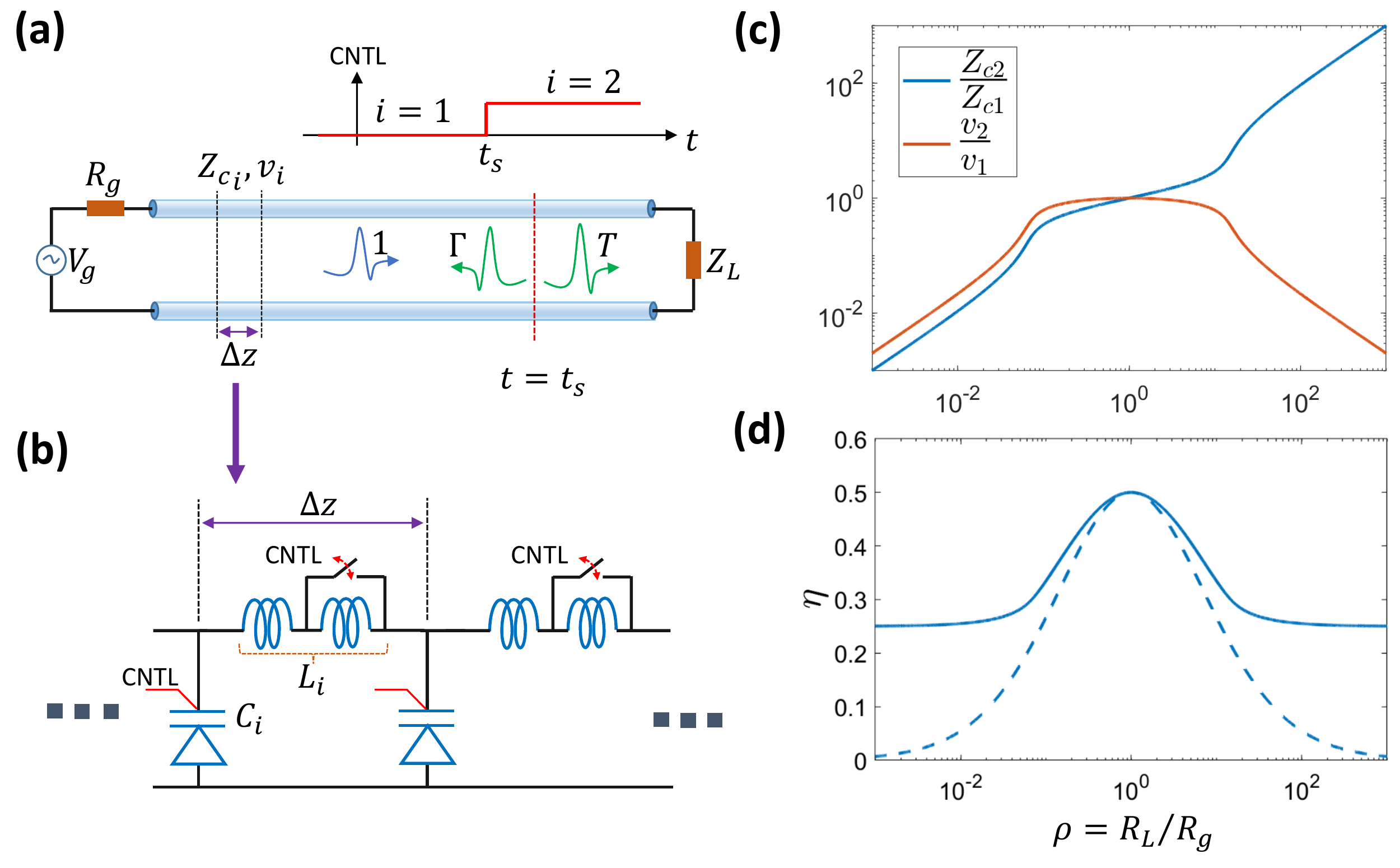}
\caption[]{(a) Physical layout of the switched TL system. (b) A possible realization of the switched TL as a discrete periodic structure of parallel varactor diodes and banks of switched series inductors.  (c) For $Z_L=R_L$. The ratio between the TL parameters required for the optimal realization versus load-generator contrast $\rho$: $Z_{c2}/Z_{c1}$ in blue line and $v_2/v_1$ in brown line. (d) Continuous line - the optimal efficiency (with switching), and dashed line - the efficiency for the
``no switching'' case.}
\label{Fig-2}
\end{figure}

Time-varying media have been suggested for various purposes and schemes such as for signal parametric amplification and delaylines \cite{Morgenthaler1958, Auld1968, Rezende1969, Felsen1970, Fante1971, Budko2009, Cervantes-Gonzalez2009, Kunz1966},  energy accumulation \cite{Lurie2016, Lurie2017, Tretyakov2018}, non-reciprocity in non-Hermitian time-Floquet systems \cite{Fleury2018}, inverse prism functionality \cite{OL_Caloz2018}, temporal-photonic crystal \cite{Halevi2016} with its real-space moving analogue \cite{Skorobogatiy2016}, unusual electromagnetic modes \cite{PRA Caloz 2018}, mixer-duplexer antenna system \cite{Caloz2017},  magnet-less non-reciprocity \cite{Sounas2013, Fleury2014, Estep2014, Hadad2016, Taravati2017, Sounas2017}, and as a means to implement synthetic magnetic field \cite{Fang2012}, as well as for wave pattern engineering \cite{Fort2016, Mattei2017}.
As opposed to previous work, here, we explore a paradigm to achieve  impedance matching for short-time pulses using temporal switching of a TL parameters.
To motivate the proposed concept, consider the system shown in Fig.~\ref{Fig-2}(a) in which a pulsed source is connected to a load through a TL. Here, $Z_c$, $R_g$, and $Z_L=R_L$, are assumed dispersion-less, and thus, real-valued. Now, assume that initially $Z_c=R_g$ so that power delivery from the source to the TL is optimal, and consider a pulse that is propagating along the TL towards the load. Since $Z_L\neq Z_c$, reflections are expected upon hitting the load. Hypothetically, however, this would be avoided if we could, while the pulse is on the TL,  switch the TL impedance to be $Z_c=Z_L$.
In this case, at the generator and at the load power delivery is optimal. Nevertheless, the  overall performance will be affected by reflections due to the temporal discontinuity.
Despite being somewhat naive, this description hints on the plausibility  of finding an optimal choice of temporal switching schemes that yield better matching performance than time-invariant matching networks. We further note that,  due to its time-variance nature,  the proposed concept is not subject to the Bode-Fano bound, and thus, opens a unique route to resolve a fundamental issue in wave-engineering with significant potential in various wave-based applications \cite{Balanis,Monticone,Bok2018}.

\emph{Pulse dynamics in abruptly switched TL}.\textemdash
We consider  the network shown in Fig.~\ref{Fig-2}(a). 
It is assumed that the TL characteristics, \emph{impedance and phase velocity}, are dispersion-less, and can be abruptly switched, at some time $t_s$, between two states $\#1$ and $\#2$, with $(Z_{ci},v_i)$, for state $\#i=1,2$. See Fig.~\ref{Fig-2}(b) for a schematic circuit realization of such a metamaterial TL, with $Z_{ci}=\sqrt{L_i/C_i}$ and $v_i=1/\sqrt{L_i C_i}$, where $L_i$ and $C_i$ are the per-unit-length distributed line inductance and capacitance (additional, and, possibly more practical, realizations are discussed in \cite[Sec.~6]{SM}).
%
%
%
%
Assume that the TL is initially at state $\#1$. The voltage on the source, $V_g(t)$, excites a forward propagating pulse with temporal width $T_1$ that propagates along the line toward the load,
\begin{equation}
V_1^+(t,z)={Z_{c1}}V_g(t-z/v_1)/\left(Z_{c1}+R_g\right).
\label{eq.300}
\end{equation}
Then, at some time $t=t_s$, that is sufficiently larger than the pulse width $T_1$, the TL characteristics are abruptly switched to state $\#2$, see Fig.~\ref{Fig-2}(a).
During the switching moment, the electric charge and the magnetic flux along the line remain continuous \cite[Sec.~1]{SM}. To satisfy this continuity, the pulse $V_1^+$ splits into a forward and backward propagating pulses, $V_2^+$ and $V_2^-$, respectively. They are related to the original pulse $V_1^+(z,t)$ via \cite{Weinstein1965, Agrawal2014, SM}
\bse
\label{eq.10}
\bea
V_2^+(z,t)&=\5T  V_1^+\left( ({v_2}/{v_1})\tau - {\zeta}/{v_1}\right ),
\label{eq.10a}
\\[1ex]
V_2^-(z,t)&=\Gamma  V_1^+\left( -({v_2}/{v_1}) \tau - {\zeta}/{v_1}\right ),
\label{eq.10b}
\eea
\ese
%
%
%
%
with $\tau=t-t_s$, $\zeta=z-z_s$ where $z_s=v_1t_s$, and the transmission and reflection coefficients are given by

\be
\label{eq.11}
\5T =\frac{1}{2}\left(\frac{v_2}{v_1}\right) \left [\frac{Z_{c2}}{Z_{c1}} +1\right ],\mbox{ }
\Gamma  =\frac{1}{2} \left(\frac{v_2}{v_1}\right) \left [\frac{Z_{c2}}{Z_{c1}} -1\right ].
\ee

Assuming that the pulse width of $V_1^+$ is $T_1$, then, it follows from Eq.~\eqref{eq.10} that the pulse width of $V_2^{\pm}$ is  $T_2=(v_1/v_2) T_1$. Thus, the TL switching from state $\#1$ to $\#2$ results in only a temporal up/down pulse compression that can be readily restored by digital or analogue means, as opposed to the complex signal distortion that is obtained by a conventional matching scheme such as in Fig.~\ref{Fig-1}(c). Furthermore, denoting $\5E_1=\|V_1^+\|^2/Z_{c1}$ and $\5E_2^\pm=\|V_2^\pm\|^2/Z_{c2}$, where $\| \ \|$ is the $\5L_2$ norm,  as the energy delivered by the corresponding pulses, the energy balance $\Delta \5E=\5E_2^+ +\5E_2^- - \5E_1$ reads,
\be
\Delta \5E = \left \{ \frac{1}{2} \left(\frac{v_\72}{v_\71}\right )\left [\frac{Z_{c2}}{Z_{c1}}+\frac{Z_{c1}}{Z_{c2}}\right ] -1 \right \} \5E_1.
\label{eq.12d}
\ee
Using Eq.~(\ref{eq.12d}) we identify three switching regimes: $(i)$ $\Delta \5E< 0$ where energy is absorbed (dissipated) from the wave-system; $(ii)$ $\Delta \5E= 0$  where no energy change occurred, and $(iii)$ $\Delta \5E>0$ where energy is pumped into the wave-system (See \cite[Sec.~2]{SM} and Fig.~S1 there). While the first two are considered `passive', the last one involves parametric amplification and is therefore considered `active'. Each of these switching cases constitutes a different wave dynamic in terms of compression, velocity, and energy conversion efficiency between source and load.

The former discussion is independent of the load type. However, for clarity, in the following we focus on a resistive, non-dispersive, load $Z_L=R_L$, whereas a general treatment of a dispersive load is given in \cite[Sec.~5]{SM}.

%
\emph{Formulation of impedance matching as a constrained optimization problem}.\textemdash
The energy delivered to the \emph{non-dispersive} resistive load $Z_L=R_L$, by the \emph{first pulse} reaching the load at $t > t_s$, reads, \cite[Sec.~3]{SM} 
\be
\5E_L=\frac{Z_{c2}/R_L}{\left ( 1+ Z_{c2}/R_L \right )^2} \left (\frac{v_2}{v_1} \right )\left [ \sqrt{\frac{Z_{c2}}{Z_{c1}}}+\sqrt{\frac{Z_{c1}}{Z_{c2}}}\right ]^2 {\5E_1},
\label{eq.22}
\ee
with $\5E_1=2(Z_{c1}/R_g)/(1+Z_{c1}/R_g)^2 \5E_{a}$ and $\5E_a=\|V_g\|^\72/2 R_g$ is the maximal available energy that the source could provide to a matched load (namely, with no switching and with $Z_0=R_L=R_g$).
%
We define the matching efficiency, in the most conservative way, as the ratio between the power that is dissipated on the load, and the maximal available power from the source $\5E_a$ plus the total power that enters to the wave network by the switching $\Delta\5E$. Thus,
\be
\eta={\5E_L}/\left[{\5E_a+\Delta \5E \mbox{H}(\Delta\5E)}\right].
\label{eq.14}
\ee
Note that in Eq.~(\ref{eq.14}) we used the Heaviside function  $\mbox{H}(x)=1$ for $x>0$ and $H(x)=0$ for $x\le0$, to imply that in the `active' case the matching efficiency accounts also for the additional power by the switching.
%
%
It should also be noted that as a quality measure an `efficiency' is not a natural parameter of a system but can be defined in various ways as to capturing (and emphasize) different and desired characteristics, see e.g. \cite{Collin}. Thus, if required, the proposed approach can be easily augmented to these definitions as well.  Inspection of Eq.~\eqref{eq.14} with Eq.~\eqref{eq.22} suggests that for a given $R_g$ and $R_L$, $\eta$  is a function of three normalized free parameters $Z_{c1}/R_g$, $Z_{c2}/R_L$ and $(v_2/v_1)$ that are constrained by a given $\Delta \5E$ of Eq.~\eqref{eq.12d} (see, e.g., Fig.~S1 \cite{SM}).
Our goal is to maximize the matching efficiency $\eta$ with respect to these free parameters. Thus, the matching problem has been replaced by a three-dimensional, generally non-convex, optimization problem. 
Note that the switching between the two states of the TL enables us to increase the number of degrees of freedom we have for the matching, from one to three, and therefore make it possible to reach optimal points that could not be reached otherwise.  This idea, may, in principle, and with more efforts, be augmented for switching between larger number of states to improve performance.

\emph{Passive switching  $[\Delta \5E=0]$}.\textemdash
By using Eq.~\eqref{eq.14} subject to Eq.~\eqref{eq.12d} with $\Delta \5E = 0$,  and setting $\rho=R_L/R_g$, $x=Z_{c1}/R_g$, $y=Z_{c2}/R_L$ and $\6v=v_2/v_1$, the efficiency optimization problem is set as, \cite[Sec.~4]{SM}
\bse
\label{eq.16}
\bea
&\max_{x,y,\6v} \left \{ 2\frac{\6v}{\rho}\frac{x^\72}{(1+x)^\72}\frac{1}{(1+y)^2 } \left [ \rho \frac{y}{x}+1\right]^\72 \right \}
\label{eq.16a}
\\[1ex]
&\text {subject to} \ \left (\rho \frac{y}{x}\right)+\left (\rho \frac{y}{x}\right)^\7{-1}=\frac{2}{\6v}.
\label{eq.16b}
\eea
\ese
In this case the optimization proposed in Eq.~\eqref{eq.16} can be  transformed to the solution of a single transcendental equation that is solved numerically \cite{SM}. Figure \ref{Fig-2}(c) depicts the required ratio of the TL characteristic parameters, before and after the switching, $Z_{c2}/Z_{c1}$ and $v_2/v_1$, as function of the load-generator contrast $\rho$, to achieve the optimal efficiency. The optimal efficiency attained by the proposed switching scheme is shown in the continuous curve in Fig.~\ref{Fig-2}(d). For comparison, the dashed curve in Fig.~\ref{Fig-2}(d) depicts the best efficiency for non switchable TL, i.e., TL with a fixed characteristic impedance $Z_c=\sqrt{R_g R_L}$.
The results in the current and following sections have been verified by circuit simulations of the
metamaterial TL that its unit cell is shown in
Fig.~\ref{Fig-2}(b), for several cases, and including non-idealities like gradual switching and the effect of TL discretization \cite[Sec.~7]{SM}.
It is clearly noted that switching of the TL parameters provides a significant increase in the trasmission efficiency in comparison with the non-switching case. Remarkably,  the increase is more dominant for the challenging high contract cases where $R_g \gg R_L$ and  $R_g \ll R_L$, and moreover, in this case becomes nearly independent on the contrast. Furthermore, returning to Fig.~\ref{Fig-1}(d), for highly dispersive load, with the switched-TL scheme, again, better results are obtained when $\rho\gg1$ or $\ll1$. Albeit somewhat logic defying, this behaviour becomes evident once noting that in this case the dynamic range for becomes larger (see Fig.~S7 in \cite{SM}).

\begin{figure}[!h]
\centering
     \includegraphics[width=85mm]{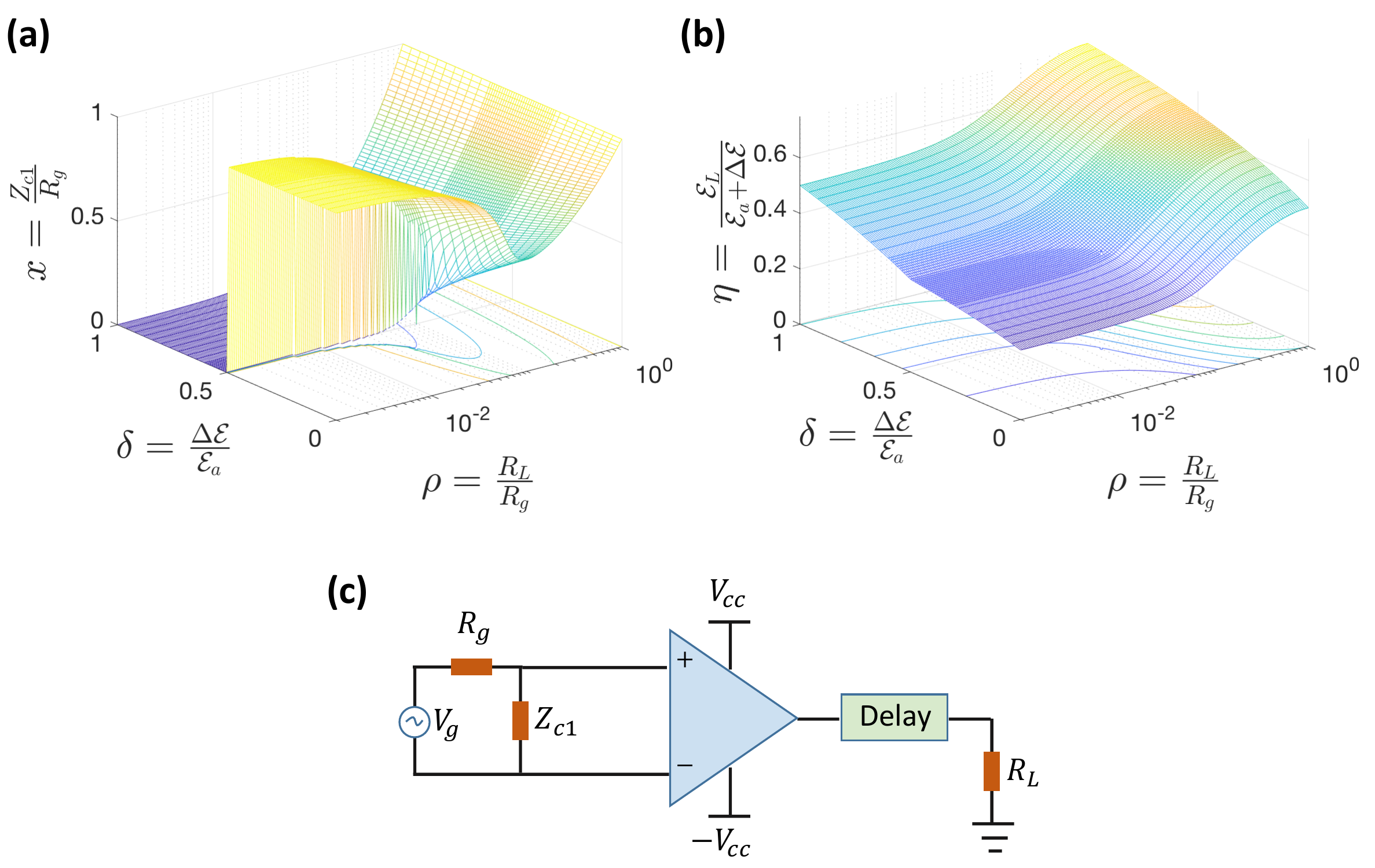}
\caption[]{(a) The optimal value of $x=Z_\7{c}/R_g$ as a function of the contrast, $\rho$, and the amount of energy balance $\delta$. (b)   The optimal value the efficiency, $\eta$ as a function of $\rho$ and $\delta$. (c) Equivalent circuit model for $0.5<\delta<1$ and high generator-load contrast $\rho$.}
\label{Fig-3}
\end{figure}

\emph{Active switching $[\Delta \5E > 0]$}.\textemdash
In the previous section signal amplification by the switching scheme was not allowed, this is due to the constraint $\Delta\5E=0$. for this reason we called this scheme `passive'. Here, conversely, we allow injection of energy into the system by the switching, i.e., $\Delta \5E >0$. With that in mind, and setting $\Delta \5E = \delta \5E_a$ with $\delta>0$, Eq.~\eqref{eq.14} becomes $\eta = \5E_L/[(1+\delta) \5E_a]$ (where $\5E_L$ also depends on $\delta$ via the characteristic impedances, see Eq.~(\ref{eq.22})). For a given energy balance between the source energy and the switching, i.e., $\delta$, The optimization problem is set as, (\cite[Sec.~4]{SM})
\bse
\label{eq.200}
\bea
&\max_{x,y,\6v} \left \{ \frac{2}{1+\delta}\frac{\6v}{\rho}\frac{x^\72}{(1+x)^\72}\frac{1}{(1+y)^2 } \left [ \rho \frac{y}{x}+1\right]^\72 \right \}
\label{eq.200a}
\\[1ex]
&\text {subject to} \ \left (\rho \frac{y}{x}\right)+\left (\rho \frac{y}{x}\right)^\7{-1}=\left [ 2+ \frac{(1+x)^2}{x} \delta \right ]\!\frac{1}{\6v},
\label{eq.200b}
\eea
where $x$, $y$, $\6v$ and $\rho$ are defined as in Eq.~\eqref{eq.16}. Note that Eq.~\eqref{eq.200} is a generalization of Eq.~\eqref{eq.16} for the case $\delta>0$.
\ese
The constraint structure in \eqref{eq.200b} is more complicated than that in Eq.~\eqref{eq.16b}, rendering an involved direct maximization of $\eta$ (in Eq.~\eqref{eq.200a}). To this end, standard optimization tools are used \cite{Comm1}. The optimization results are shown in Fig.~\ref{Fig-3} as function of the contrast $\rho=R_L/R_g$ and the additional energy to the wave system by the switching $\delta=\Delta\5E/\5E_a$. In panels (a) and (b) we show $x=Z_{c1}/R_g$  and $\eta$, respectively. Similar figures  for $y=Z_{c2}/R_L$ and $\6v$ are shown in Fig.~S2 \cite{SM}.
We note that in the `passive' matching case, namely, for $\Delta\5E=0$ ($\delta=0$), the optimization procedure leads to a single optimum point, which is, therefore, a global maximum.
As opposed to that, for $\delta>0$, there are several local optima. Nevertheless, the global optimum behaves similar to the `passive', $\delta=0$, solution.
However, as soon as $\delta$ increases above the $0.5$ threshold, namely, as soon as the power that enters into the wave-system by the switching mechanism exceeds the maximal power that can be delivered to a matched load without switching (recall that for $R_g=R_L=Z_0$, $\5E_L=0.5\5E_a$), the optimization, and therefore, also the matching dynamics significantly alter for the high contrast cases, see e.g., Fig.~\ref{Fig-3}(a) for $\rho\lesssim0.1$.
In this case, interestingly, the global optimum exhibits a remarkably different matching mechanism compare to the case of $0<\delta<0.5$, as evident by the `jump' in Fig.~\ref{Fig-3}(a). Here, the global optimum is obtained when the line impedance at state $\#1$ is extremely low, i.e., $Z_{c1}/R_g\ll1$. Thus, the line merely sense the generator signal. Then, at the switching moment to state $\#2$, the sampled signal is amplified by the power that enters through the switching process. Hence, in these cases, the actual role of $V_1^+$ is to provide a sampling of the source's waveform that would be amplified later by the switching to provide the energy for the matching. In this regime, therefore, the switched matching system can be described as a cascaded system of a sampler and an ideal amplifier that buffers between the generator and the load, as schematically shown in Fig.~\ref{Fig-3}(c).
In light of the nature of this, so-called, `active' regime, here, the matching efficiency, as defined in Eq.~(\ref{eq.14}) can exceed values up to $\eta=0.75$ for $\delta=1$ as shown in Fig.~\ref{Fig-3}(c), and up to $\eta=1$ for $\delta\gg1$ (\cite[Sec.~4.2]{SM}).

\emph{Conclusions and Discussion}.\textemdash
In this letter we have discussed  a  paradigm for short-pulse impedance matching in TL networks by switching the TL's parameters between two states.  Generally, this matching system may be passive or active depending on the two switching states. The latter are determined by solving a nonlinear constrained optimization for the matching efficiency.  We have demonstrated that the proposed approach can overcome major challenges that exist in other matching techniques, particularly for cases of high contrast between the source and load impedances, and with high load dispersion,  and moreover it enables minimal-distortion and efficient wideband signal transmission. %
Our approach, therefore,  opens possible venues to cope with today's ever-ending demand for high-speed ultra-wideband communication, for small antennas matching, it may be applied to achieve wideband cloaking, and more. In addition, we note that a byproduct of the suggested switching scheme is the ability to obtain significant delay of signals along the line. Thus, an alternative design goal can be taken by defining a different quality measure where the switched TL is considered as a `controlled delay line' to be used in, e.g., true time delay pulsed systems \cite{TrueTimeDelay1,TrueTimeDelay2}. Lastly, we note that the proposed paradigm is suitable only for finite time-duration signals; using time-variant matching  for time-harmonic signals is yet an open question.

\begin{acknowledgments}
A.~Shlivinski would like to acknowledge a partial support by the Israel Science Foundation (grant No.~545/15).
Y.~Hadad would like to acknowledge support by the Alon Fellowship granted by the Israeli
Council of Higher Education (2018-2021).
\end{acknowledgments}


\begin{thebibliography}{1}


\bibitem{Collin}
R. E. Collin, \emph{Foundations for Microwave Engineering}, McGraw-Hill, 1966.

\bibitem{Kinsler}
L. E. Kinsler, and A. R. Frey, \emph{Fundamentals of Acoustics}, John Wiley \& Sons, 2nd ed. 1962.


\bibitem{Bode}
H. W. Bode, \emph{Network Analysis and Feedback Amplifier Design}, Van Nostrand, New York, 1945.


\bibitem{Fano}
R. M. Fano, ``Theoretical limitations on the broadband matching of arbitrary impedances'', Technical Report NO. 41, Research Laboratory of Electronics, MIT, 1948.


\bibitem{Acher2009}
O. Acher, J. M. L. Bernard, P. Maréchal, A. Bardaine, F. Levassorta, ``Fundamental constraints on the
performance of broadband ultrasonic matching structures and absorbers,''
\emph{The Journal of the Acoustical Society of America} {\bf125}, pp.~1995-2005 (2009).

\bibitem{Pozar}
D. M. Pozar, \emph{Microwave Engineering}, John Wiely \& Sons, 2nd ed. 1998 (p. 295-297),


\bibitem{Monticone}
F. Monticone, A. Al\'{u}, ``Invisibility exposed: physical bounds on passive cloaking'',
\emph{Optica} {\bf3} (7), 718-724 (2016).


\bibitem{Balanis}
C. A. Balanis, \emph{Modern Antenna Handbook}, John Wiley \& Sons, (2008), Ch. 2.

\bibitem{PaiYen}
P.-Y. Chen, J. Soric, A. Al\'{u}, ``Invisibility and cloaking based on scattering cancellation,'' \emph{Advanced Materials} {\bf24} (44) pp.281-304 (2012).



\bibitem{DELANEY}
B.  Delaney, ``Wideband Radar,'' \emph{Lincoln Laboratory Journal}, {\bf18}, 2, pp. 88-89, (2010)


\bibitem{Bok2018}
E. Bok, Jong J. Park, H. Choi, Chung K. Han, Oliver B. Wright, and Sam H. Lee, ``Metasurface for Water-to-Air Sound Transmission,''
\emph{Phys. Rev. Lett.}, {\bf 120}, 044302 (2018).



\bibitem{SM}
Supplementary Material available online.

\bibitem{Comm2}
Note that for non-dispersive load, $Z_L=R_L$, for maximal transmission efficiency, $Z_{c,\mbox{\scriptsize opt}}=\sqrt{R_gR_L}$, with optimum obtained when $R_L/R_g=1$.






\bibitem{Sussman-Fort}
S. E. Sussman-Fort, R. M. Rudish, ``Non-Foster Impedance Matching of Electrically-Small Antennas,'' IEEE Trans. on Ant.  Prop., {\bf57} 8, pp. 2230-2241, (2009).

\bibitem{Pai-Yen}
P. -Yen Chen, C. Argyropoulos, and A. Al\'{u}, ``Broadening the Cloaking Bandwidth with Non-Foster Metasurfaces,'' \emph{Phys. Rev. Lett.}, {\bf 111}, 233001 (2013).


\bibitem{Soric}
J. C. Soric, A. Al\'{u},  ``Wideband tunable and non-Foster mantle cloaks,'' Radio Science Meeting (USNC-URSI NRSM), 2014 United States National Committee of URSI National.


\bibitem{Stearns2011}
S. D. Stearns, ``Non-foster circuits and stability theory,'' in the proceedings of the International Symposium on Antennas and Propagation (APSURSI), 2011 IEEE.

\bibitem{Stearns2012}
S. D. Stearns, ``Incorrect stability criteria for non-foster circuits,'' in the proceedings of Antennas and Propagation Society International Symposium (APSURSI), 2012 IEEE.


\bibitem{Morgenthaler1958}
F. R. Morgenthaler, ``Velocity modulation of electromagnetic waves,'' \emph{IRE Trans. Microwave Theory and Techniques}, {\bf 6}, pp. 167-172 (1958).


\bibitem{Auld1968}
B. A. Auld, J. H. Collins, and H. R. Zapp, ``Signal processing in a non-periodically time-varying magnetoelastic medium,'' \emph{Proc. IEEE} {\bf 56}, pp. 258-272 (1968).



\bibitem{Rezende1969}
S. M. Rezende and F. R. Morgenthaler, ``Magnetoelastic waves in time-varying magnetic field I: Theory,'' \emph{J. Appl. Phys.} {\bf 40}, pp. 524-536 (1969).



\bibitem{Felsen1970}
L. B. Felsen, G. M. Whitman, ``Wave propagation in time-varying media,'' \emph{IEEE Trans. Ant. Prop.} {\bf 18}, 2, pp. 242-253 (1970).


\bibitem{Fante1971}
R. L. Fante, ``Transmission of electromagnetic Waves into time-varying media,'' \emph{IEEE Ant. Prop.}, {\bf 19}, No. 3 pp. 417-424 (1971).


\bibitem{Budko2009}
Neil V. Budko, ``Electromagnetic radiation in a time-varying background medium,'' \emph{Phys. Rev. A}, {\bf80}, 053817 (2009).


\bibitem{Cervantes-Gonzalez2009}
J. R. Zurita-Sanchez, P. Halevi, and J. C. Cervantes-Gonzalez, ``Reflection and transmission of a wave incident on a slab with a time-periodic dielectric function $\epsilon(t)$,'' \emph{Phys. Rev. A}, {\bf79}, 053821 (2009).


\bibitem{Kunz1966}
D. E. Holberg, and K. S. Kunz, ``Parametric properties of fields in a slab of time-varying permittivity,'' \emph{IEEE Trans. Ant. Prop.}, {\bf14}, No. 2, pp.183-194 (1966).


\bibitem{Lurie2016}
Konstantin A. Lurie and Vadim V. Yakovlev, ``Energy Accumulation in Waves Propagating in Space- and Time-Varying Transmission Lines,'' \emph{IEEE Antennas and Wireless Propagation Letters} {\bf15} pp. 1681-1684 (2016).

\bibitem{Lurie2017}
K. A. Lurie, D. Onofrei, W. C. Sanguinet, S. L. Weekes, and Vadim V. Yakovlev, ``Energy Accumulation in a Functionally Graded Spatial-Temporal Checkerboard,'' \emph{IEEE Antennas and Wireless Propagation Letters}{\bf16} pp. 1496-1499 (2017).

\bibitem{Tretyakov2018}
M. S. Mirmoosa, G. A. Ptitcyn, V. S. Asadchy and S. A. Tretyakov, ``Unlimited Accumulation of Electromagnetic Energy Using Time-Varying Reactive Elements,'' \emph{arXiv:1802.07719v1} (2018).

\bibitem{Fleury2018}
T. T. Koutserimpas and R. Fleury, ``Nonreciprocal Gain in Non-Hermitian Time-Floquet Systems,'' \emph{Phys. Rev. Lett.}, {\bf120}, 087401 (2018).



\bibitem{OL_Caloz2018}
A. Akbarzadeh, N. Chamanara, and C. Caloz, ``Inverse prism based on temporal discontinuity and spatial dispersion,'' \emph{arXiv:1710.11264} (2017).



\bibitem{Halevi2016}
Juan Sabino Martınez-Romero, O. M. Becerra-Fuentes, and P. Halevi, ``Temporal photonic crystals with modulations of both permittivity and permeability,'' \emph{Phys. Rev. A}, {\bf93}, 063813 (2016).


\bibitem{Skorobogatiy2016}
H. Qu, Zo\'{e}-L. Deck-L\'{e}ger, C. Caloz, and M. Skorobogatiy, ``Frequency generation in moving photonic crystals,'' \emph{J. Opt. Soc. Am. B,} {\bf 33}, pp. 1616-1626 (2016).

\bibitem{PRA Caloz 2018}
N. Chamanara, Zo\'{e}-L. Deck-L\'{e}ger, C. Caloz, and D. Kalluri, ``Unusual electromagnetic modes in space-time-modulated dispersion-engineered media,'' \emph{Phys. Rev. A,} {\bf 97}, 063829 (2018).  

\bibitem{Caloz2017}
S. Taravati and C. Caloz, ``Mixer-Duplexer-Antenna Leaky-Wave System Based on Periodic Space-Time Modulation,'' \emph{IEEE Tran. Ant. Prop.}, {\bf 65}, No. 2, pp. 442-452 (2017). 


\bibitem{Sounas2013}
D. L. Sounas, C. Caloz, A. Al\'{u}, ``Giant non-reciprocity at the subwavelength scale using angular momentum-biased metamaterials,'' \emph{Nat. comm.} {\bf4}, 2407 (2013).


\bibitem{Fleury2014}
R. Fleury, D. L. Sounas, C. F. Sieck, M. R. Haberman, A. Al\'{u}, ``Sound isolation and giant linear nonreciprocity in a compact acoustic circulator,'' \emph{Science} {\bf343} (6170), pp. 516-519	(2014).

\bibitem{Estep2014}
N. A. Estep, D. L. Sounas, J. C. Soric, A. Al\'{u}, ``Magnetic-free non-reciprocity and isolation based on parametrically modulated coupled-resonator loops,'' \emph{Nat. Phys.} {\bf10} (12), pp. 923-926	(2014).


\bibitem{Hadad2016}
Y. Hadad, J. C. Soric, A. Al\'{u}, ``Breaking temporal symmetries for emission and absorption,''
\emph{Proceedings of the National Academy of Sciences} {\bf113} (13), 3471-3475 (2016).


\bibitem{Taravati2017}
S. Taravati, N. Chamanara, and C. Caloz, ``Nonreciprocal electromagnetic scattering from a periodically space-time modulated slab and application to a quasisonic isolator,'' \emph{Phys. Rev. B} {\bf 96}, (16) (2017).

\bibitem{Sounas2017}
D. L. Sounas, A. Al\'{u}, ``Non-reciprocal photonics based on time modulation,'' \emph{Nat. Phot.} {\bf11} (12), 774 (2017).


\bibitem{Fang2012}
K. Fang, Z. Yu, S. Fan, ``Realizing effective magnetic field for photons by controlling the phase of dynamic modulation,'' \emph{Nat. phot.} {\bf6} (11), 782 (2012).


\bibitem{Fort2016}
V. Bacot, M. Labousse, A. Eddi, M. Fink, and E. Fort, ``Time reversal and holography with
spacetime transformations,'' \emph{Nat. Phys.}, {\bf12}, pp. 972-977 (2016).


\bibitem{Mattei2017}
G. W. Milton, and  O. Mattei, ``Field patterns: a new mathematical object,'' \emph{Proc. R. Soc. A} {\bf473}:20160819 (2017).

 
\bibitem{Weinstein1965}
H. Weinstein, ``Linear Signal Stretching in a Time-Variant System,'' \emph{IEEE Transactions on Circuit Theory} {\bf12} pp. 157-164 (1965).

\bibitem{Agrawal2014}
Y. Xiao, D. N. Maywar, and G. P. Agrawal, ``Reflection and transmission of electromagnetic waves at a temporal boundary,'' Opt. Lett. {\bf39}, No. 3 (2014).



\bibitem{Comm1}
Nevertheless, the case of $R_L=R_g$ ($\rho=1$) can be treated heuristically as discussed in \cite{SM} to give $x=y=1$, $\6v=1+2\delta$ and $\eta=(1+2\delta)/(2+2\delta)$ to give $\eta=0.75$ at $\delta=1$ as can also be noted in Fig.~\ref{Fig-3}.


\bibitem{TrueTimeDelay1}
Betty L. Anderson and Craig D. Liddle, ``Optical true time delay for phased-array antennas: demonstration of a quadratic White cell,'' \emph{Optics Letters},  {\bf41}, (23), pp. 4912-4921 (2002).

\bibitem{TrueTimeDelay2}
R. Rotman, M. Tur, and L. Yaron,``True Time Delay in Phased Arrays,''\emph{Proceedings of the IEEE}, {\bf104}, (3), pp. 504-518 (2016).





%
%
%
%
%

\end{thebibliography}
\end{document}